\documentclass[a4paper,12pt]{article}
\usepackage{epsfig}
\newcommand{\com}[1]{}
\renewcommand{\vec}[1]{\relax\ifmmode\mathchoice
{\mbox{\boldmath$\relax\displaystyle#1$}}
{\mbox{\boldmath$\relax\textstyle#1$}}
{\mbox{\boldmath$\relax\scriptstyle#1$}}
{\mbox{\boldmath$\relax\scriptscriptstyle#1$}}\else
\hbox{\boldmath$\relax\textstyle#1$}\fi}
\begin{document}
\pagestyle{myheadings}
\markboth{Helbing/Sch\"onhof/Kern: 
Volatile Decision Dynamics}
{Helbing/Sch\"onhof/Kern: 
Volatile Decision Dynamics}
\title{\mbox{}\\[-2.2cm]{\Large\textbf{Volatile Decision Dynamics:\protect\linebreak}
{\large\bf Experiments, Stochastic Description,\protect\linebreak 
Intermittency Control, and Traffic Optimization}}\\[-0cm]}
\author{\large Dirk Helbing,$^{1, 2, 3}$ Martin Sch\"onhof,$^1$ and Daniel 
Kern$^1$\\[4mm]
\normalsize $^1$ Institute for Economics and Traffic, Dresden University of
Technology,\\
\normalsize D-01062 Dresden, Germany, {\tt helbing@trafficforum.org, www.helbing.org}\\[2mm]
\normalsize $^2$ Collegium Budapest---Institute for Advanced Study,\\
\normalsize Szenth\'{a}roms\'{a}g u. 2, H-1014 Budapest, Hungary\\[2mm]
\normalsize $^3$ CCM---Centro de Ci\^{e}ncias Matem\'{a}ticas, Universidade da Madeira,\\
\normalsize Campus Universit\'{a}rio da Penteada, Pt-9000-390 Funchal, Madeira, Portugal}
\maketitle
{\bf The coordinated and efficient distribution of limited resources by individual
decisions is a fundamental, unsolved problem. 
When individuals compete for road capacities, time, space, money, goods, etc., 
they normally make decisions based on aggregate rather than complete
information, such as TV news or stock market indices. 
In related experiments, we have observed
a volatile decision dynamics and far-from-optimal payoff distributions.
We have also identified ways of information presentation that can 
considerably improve the overall performance of the system. In order to determine 
optimal strategies of decision guidance by means of user-specific recommendations,
a stochastic behavioural description is developed.  These strategies manage
to increase the adaptibility to changing conditions
and to reduce the deviation from the time-dependent user equilibrium, 
thereby enhancing the average and individual payoffs. 
Hence, our guidance strategies can increase the performance of all users by reducing 
overreaction and stabilizing the decision dynamics. These results are 
highly significant for predicting decision behaviour, for reaching 
optimal behavioural distributions by decision support
systems, and for information service providers. One of the promising fields of
application is traffic optimization.}

\section{Introduction}

Optimal route guidance strategies in overloaded 
traffic networks, for example, require reliable traffic forecasts
(see Fig.~\ref{Fig1}). These are extremely challenging
for two reasons: First of all, traffic dynamics is very complex. However, after more than 50 years of 
traffic research, physicists have recently gained at least a semi-quantitative understanding  
of it based on the concept of self-driven, non-linearly interacting many-particle systems \cite{REVIEW}. 
The second and more serious problem is the invalidation of forecasts by the
driver reactions to route choice recommendations. 
Nevertheless, some keen scientists hope
to solve this long-standing problem by means of an iteration scheme 
\cite{SchSel,Ramming,BarDin98,Wahle,IEE,BenPaKa91,MahJa91,ArnPaLi91,Adler}:
If the driver reaction was known from experiments 
\cite{MahJou,BonFir,Iida,Chen,Kuehne,Hall,Kha,Selten,KouPoBen95,MahSt88,Bonsall}, 
the resulting traffic situation could be calculated,
yielding improved route choice recommendations, etc. Given this iteration scheme 
converges, it would facilitate optimal recommendations and reliable
traffic forecasts anticipating the driver reactions. 
Based on empirically determined transition and compliance probabilities, we will
develop a new procedure in the following, which would even
allow us to reach the optimal traffic distribution
in one single step and in harmony with the forecast.
\par\begin{figure}[htbp]
\begin{center}
\end{center}
\caption[]{Schematic illustration of a day-to-day route choice scenario. Each day,
the drivers have to decide between two alternative routes, 1 and 2. Note that, due to the
different number of lanes, route 1 has a higher capacity than route 2. The latter is, therefore,
used by less cars.\label{Fig1}}
\end{figure}
The solution of this difficult and practically relevant problem
requires several concepts and methods from physics. First of all, we will identify the
essential role and reason of {\em intermittency} in scenarios with repeated decisions 
(see Sec.~\ref{Volatile}). Second, we will derive a {\em non-linear feedback mechanism}
for {\em intermittency control} (see Sec.~\ref{Control}). In addition, 
we will develop a stochastic description of the decision behavior (see Sec.~\ref{MASTEReq})
and evaluate the corresponding transition and compliance probabilities
including their time-dependence (see Sec.~\ref{Control}).

\section{Experimental setup and previous results}
\label{Setup}
\begin{figure}[htbp]
\begin{center}
\end{center}
\caption[]{Schematic illustration of the decision experiment. Several test persons
have to take decisions based on the aggregate information their computer displays. The 
computers are connected and can, therefore, exchange information. However,
a direct communication among players is suppressed.
\label{Fig2}}
\end{figure}
To determine the route choice behavior, Schreckenberg, Selten {\em et al.} \cite{Selten} 
have recently carried out a decision experiment \cite{beheco}
(see Fig.~\ref{Fig2}). $N$ test persons had to repeatedly decide between 
two alternatives  $1$ and $2$ (the routes) and should try to maximize their resulting payoffs
(describing something like the speeds or inverse travel times). To reflect the competition for 
a limited resource (the road capacity), the received payoffs
\begin{equation} 
P_1(n_1) = P_1^0 - P_1^1 n_1 \quad \mbox{and}  \quad P_2(n_2) = P_2^0 - P_2^1 n_2
\end{equation} 
went down with the numbers of test persons $n_1$ and $n_2 = N- n_1$ deciding for
alternatives 1 and 2, respectively.  The {\em user 
equilibrium} corresponding to equal payoffs for both
alternative decisions is found for a fraction 
\begin{equation}
 f_1^{\rm eq} = \frac{n_1}{N} = \frac{P_2^1}{P_1^1+P_2^1} + 
 \frac{1}{N} \frac{P_1^0-P_2^0}{P_1^1+P_2^1} 
\end{equation} 
of persons choosing alternative 1.
The {\em system optimum} corresponds to the maximum of
the total payoff $n_1 P_1(n_1) + n_2P_2(n_2)$, which lies by an amount of
\begin{equation}
   \frac{1}{2N}\frac{P_1^0-P_2^0}{P_1^1+P_2^1} 
\end{equation} 
below the user optimum. Therefore, only
experiments with a few players allow to find out, whether the test persons adapt to the
user or the system optimum. Small groups are also more suitable 
for the experimental investigation of the fluctuations in the system
and of the long-term adaptation behavior. 
Schreckenberg, Selten {\em et al.} found that, on average, the test groups adapted
relatively well to the user equilibrium. However, 
although it appears reasonable to stick to the same 
decision once the  equilibrium is reached, the standard deviation stayed at a finite level. 
This was not only observed in {\em ``treatment'' 1}, where all  players knew only their own 
(previously experienced) payoff, but 
also in {\em treatment 2}, where
the payoffs $P_1(n_1)$ and $P_2(n_2)$ for both, 1- {\em and} 2-decisions, 
were transmitted to all players  (analogous to radio news). 
Nevertheless, treatment 2 could decrease the changing rate
and increase the average payoffs. For details regarding the 
statistical analysis see Ref.~\cite{Selten}.
\par\unitlength0.85cm
\begin{figure}[htbp]
\begin{center}
\begin{picture}(14,13.5)(-0.2,3.5)
\put(-0.35,16.9){\includegraphics[height=13.8\unitlength, angle=-90]{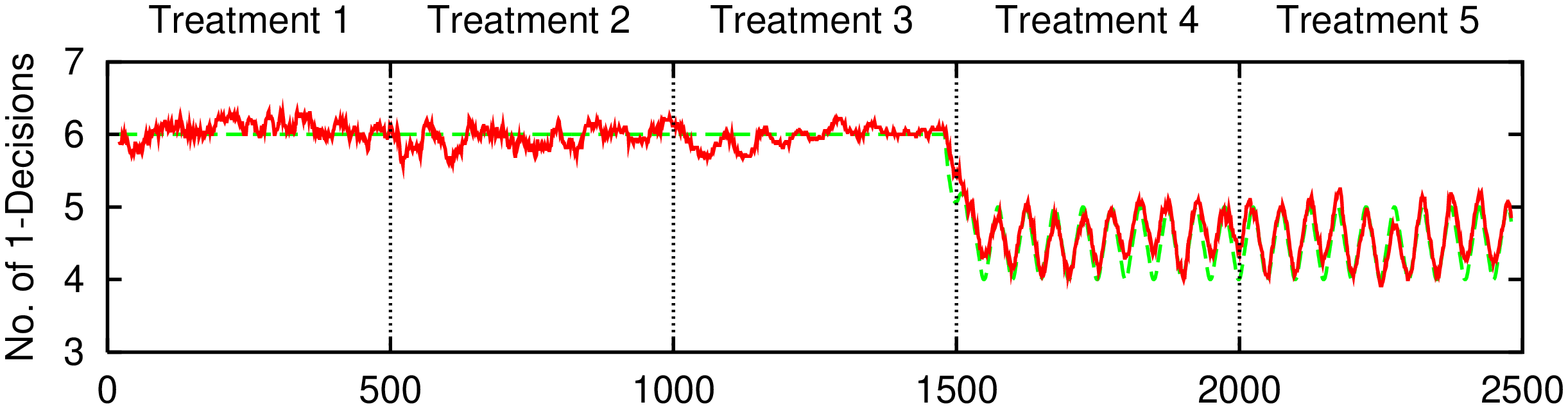}}
\put(-0.35,13.6){\includegraphics[height=13.8\unitlength, angle=-90]{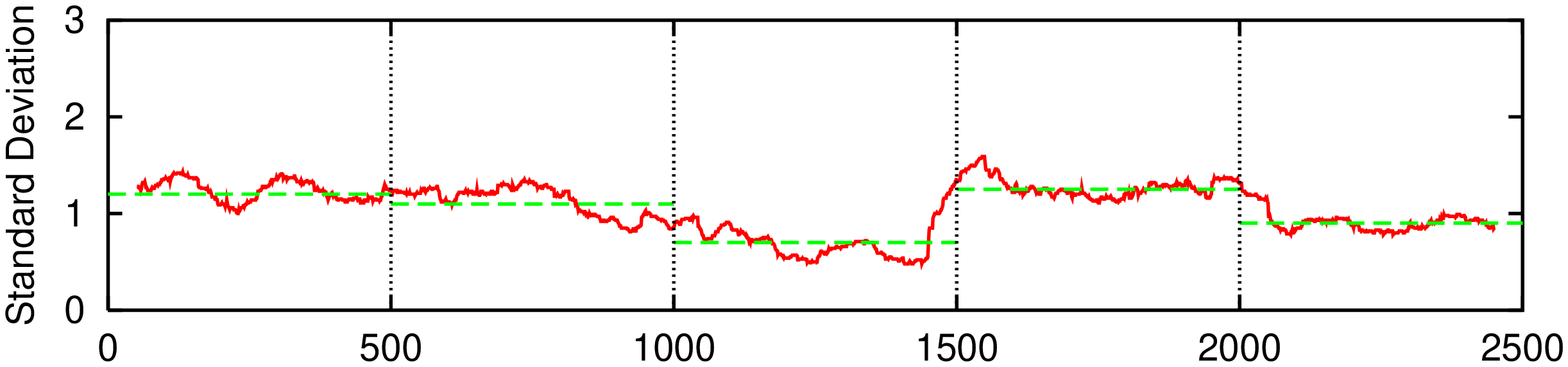}}
\put(-0.35,10.4){\includegraphics[height=13.8\unitlength, angle=-90]{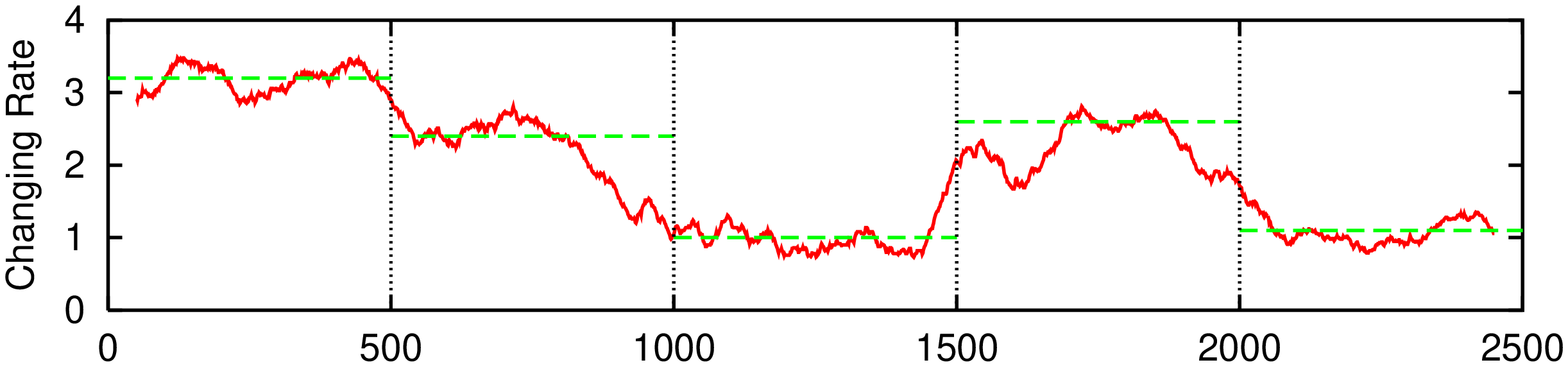}}
\put(-0.35,7.2){\includegraphics[height=13.8\unitlength, angle=-90]{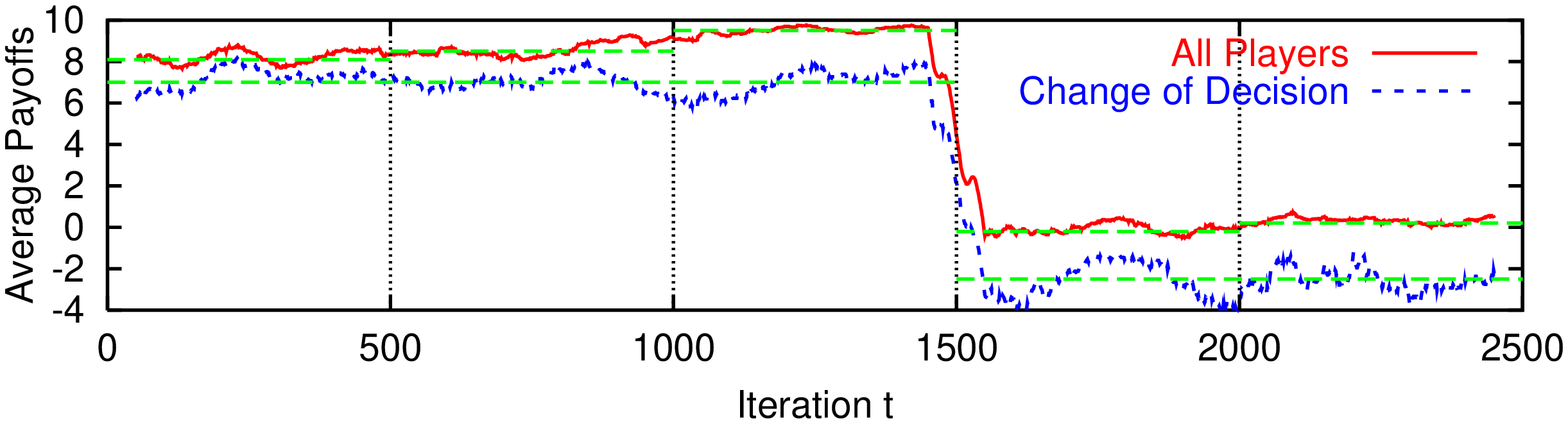}}
\put(1.1,14.05){\sf\textbf{a}}
\put(1.1,10.9){\sf\textbf{b}} 
\put(1.1,7.7){\sf\textbf{c}} 
\put(1.1,4.5){\sf\textbf{d}} 
\end{picture}
\end{center}
\caption[]{Overview of treatments 1 to 5 
(with $N=9$ and payoff parameters $P_2^0=28$,
$P_1^1 = 4$, $P_2^1 = 6$, and $P_1^0 = 34$ for $0 \le t \le 1500$, but a zick-zack-like
variation between $P_1^0 = 44$ and $P_1^0 = -6$ with a period of $50$ for $1501 \le t \le 2500$):
{\sf\textbf{(a)}} Average number of decisions for
alternative 1 (solid line) compared to the user equilibrium (dashed line), 
{\sf\textbf{(b)}} standard deviation of the number of 1-decisions from the user equilibrium,
{\sf\textbf{(c)}} number of decision changes from one iteration to the next one,
{\sf\textbf{(d)}} average payoff per iteration
for players who have changed their decision and for all players. The latter increased
with a reduction in the changing rate, but normally stayed below the 
payoff in the user equilibrium (which is 1 on average in treatments 4 and 5, otherwise 10). 
The displayed moving time-averages
[{\sf {(a)}} over 40 iterations, {\sf {(b)-(d)}} over 100 iterations] illustrate the systematic
response to changes in the treatment every 500 iterations. Dashed lines in {\sf {(b)-(d)}} 
show estimates of the stationary values after the transient period (to guide the eyes), 
while time periods around  the dotted lines are not significant. Compared to treatment 1,
treatment 3 managed to reduce the changing rate 
and to increase the average payoffs 
(three times more than treatment 2 did). 
These changes were systematic for {\it all} players (see Fig.~\ref{Fig4}). 
In treatment 4, the changing rate and the standard deviation went up,
since the user equilibrium changed in time. The user-specific recommendations in treatment 5 
could almost fully compensate for this. 
The above conclusions
are also supported by additional experiments with single treatments.\label{Fig3}}
\end{figure}
To explain the mysterious persistence in the changing behavior and explore
possibilities to suppress it, we have repeated these experiments with more iterations and 
tested additional treatments. In the beginning,  all treatments were consecutively
applied to the same players in order to determine the response to different kinds of 
information (see Fig.~\ref{Fig3}). Afterwards, single treatments and variants of them
have been repeatedly tested with different players to check our conclusions. 
Apart from this, we have generalized the experimental setup in the sense that it
was not anymore restricted to route choice decisions: The test persons did not have any
idea of the payoff functions in the beginning, but had to develop their own
hypothesis about them. In particular, the players did not know that the payoff decreased
with the number of persons deciding for the same alternative.
\par
\begin{figure}[htbp]
\unitlength1.4cm
\begin{center}
\begin{picture}(6,12)
\put(0,12){\includegraphics[width=4\unitlength, angle=-90]{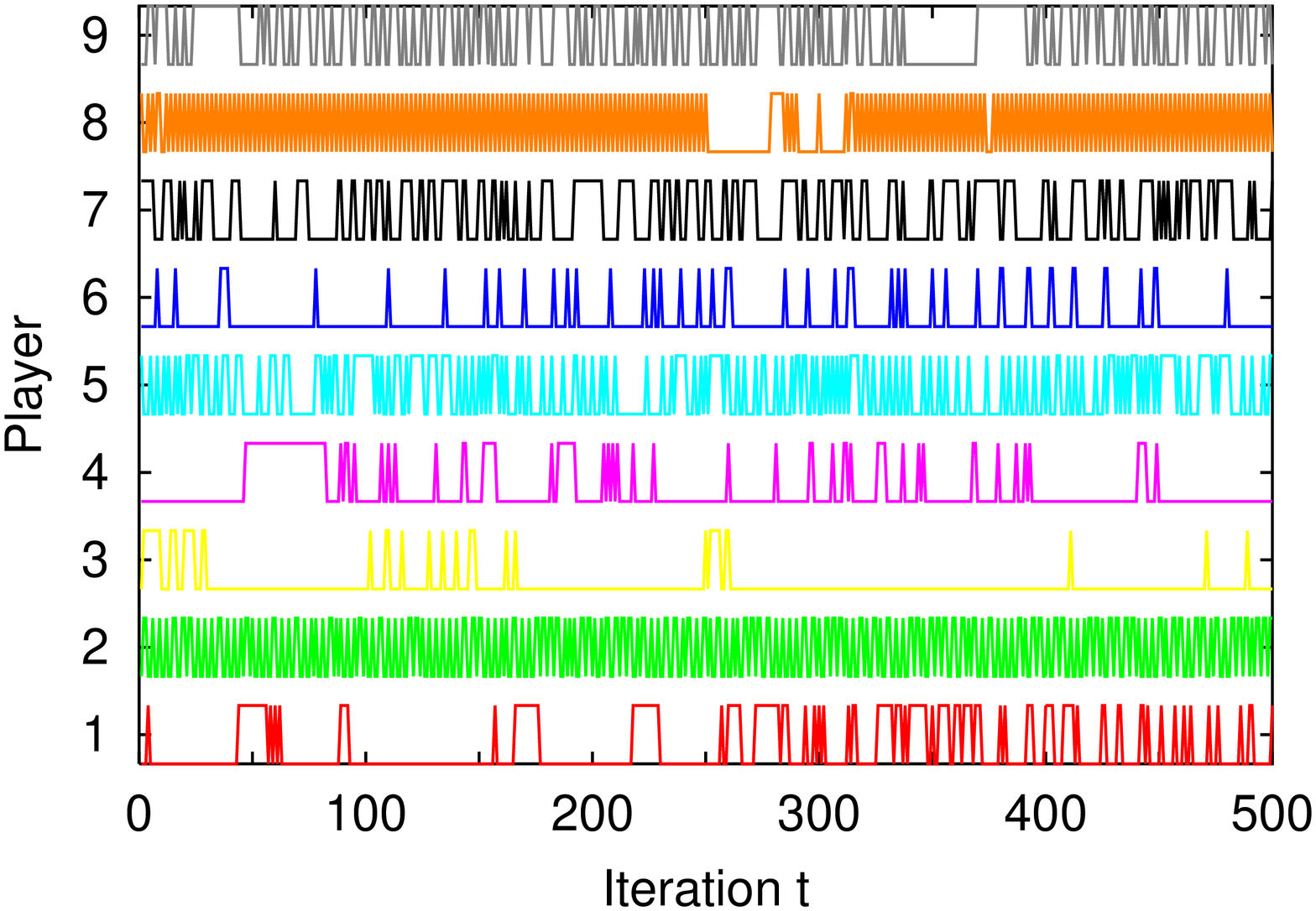}}
\put(0,8){\includegraphics[width=4\unitlength, angle=-90]{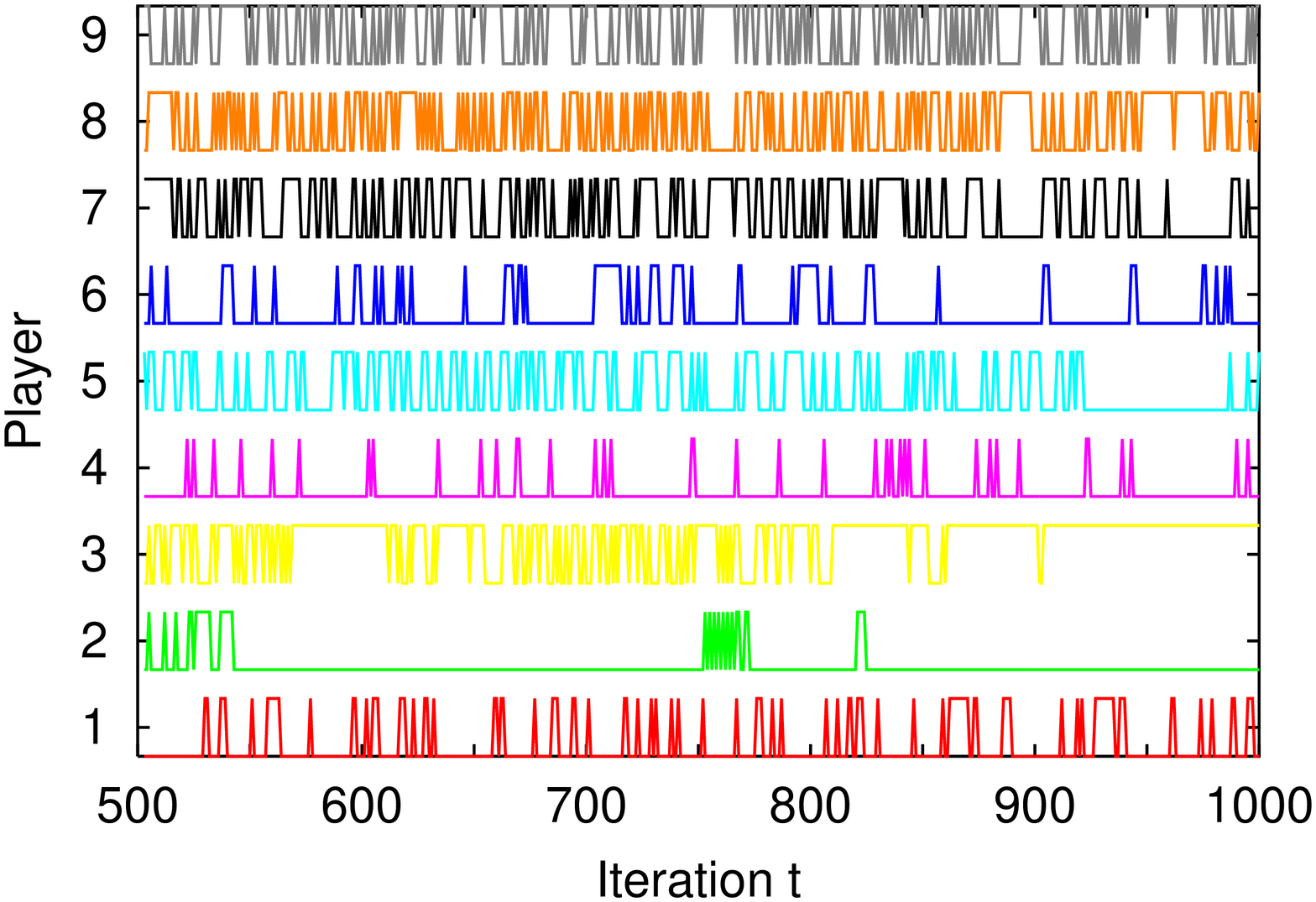}}
\put(0,4){\includegraphics[width=4\unitlength, angle=-90]{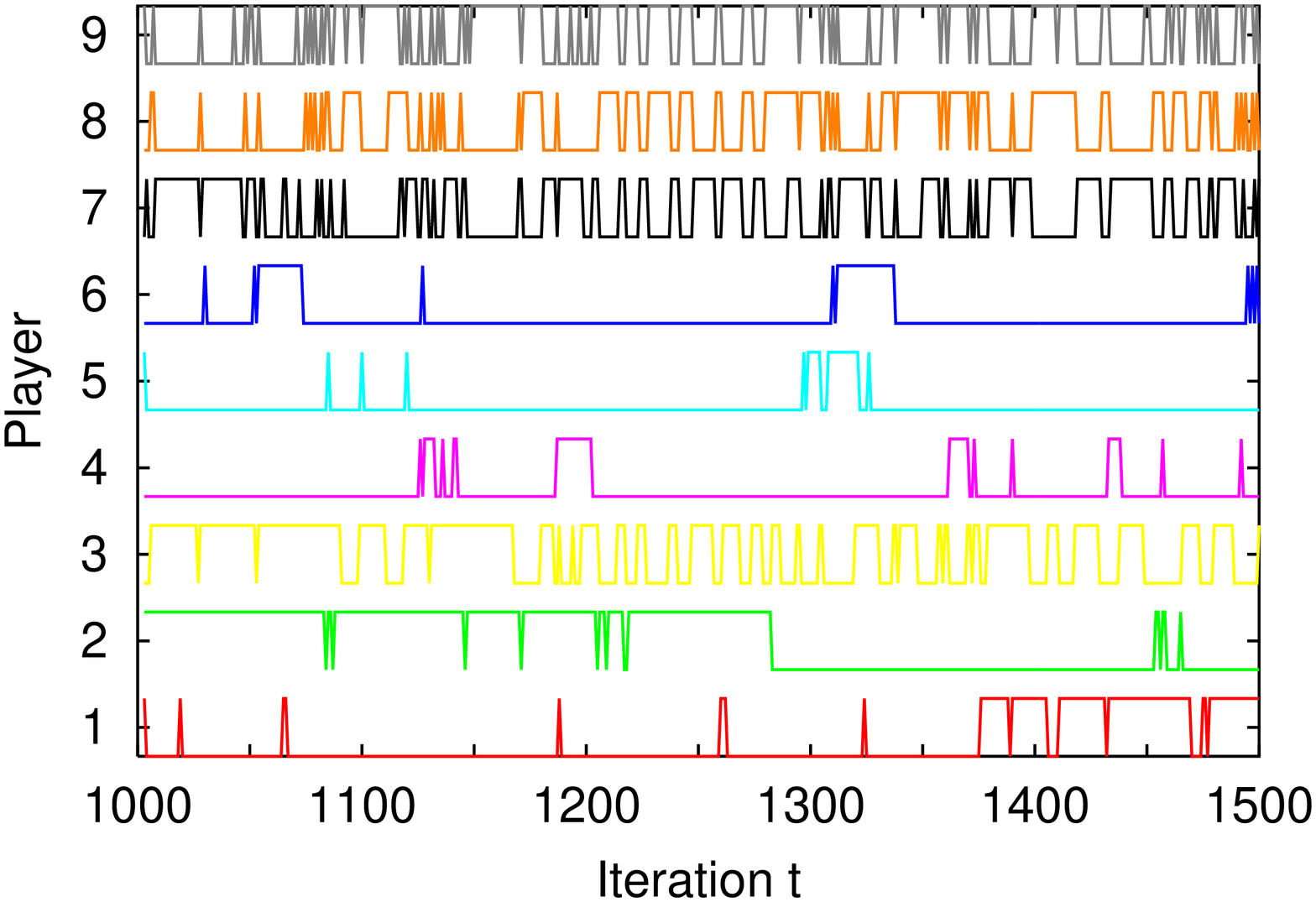}} 
\put(0.1,11.6){\sf\textbf{a}} 
\put(0.1,7.6){\sf\textbf{b}} 
\put(0.1,3.6){\sf\textbf{c}} 
\end{picture}
\end{center}
\caption[]{Comparison of the individual decision behaviors under {\sf\textbf{(a)}}
treatment 1, {\sf\textbf{(b)}} treatment 2, and {\sf\textbf{(c)}} 
treatment 3.
The upper values correspond to a decision for alternative 2, the lower ones for
alternative 1. Note that some test persons showed similar behaviors 
(either more or less the same or almost opposite ones), 
although they could not talk to each other. This
shows that there are some typical strategies how to react to specific information.
The group has, in fact, to develop
complementary strategies in order to reach a good adaptation performance. Identical
strategies would perform poorly (as in the minority game \cite{Arthur,minor1,minor2,minor3}). 
Despite the mentioned complementary behavior,
there is a characteristic reaction to changes in the treatment. For example,
compared to treatment 2 all players reduce their changing rate in treatment 3.
\label{Fig4}}
\end{figure}
In {\em treatment 3}, every test person was informed about
the own payoff $P_1(n_1)$ [or $P_2(n_2)$] {and} the {\em potential payoff}
\begin{equation}
 P_2(N-n_1+\epsilon N)= P_2(n_2) - \epsilon N P_2^1
\end{equation} 
[or $P_1(N-n_2+\epsilon N)  = P_1(n_1) - \epsilon N P_1^1$] 
he or she would have obtained, if a fraction $\epsilon$ of persons 
had additionally chosen the other alternative  (here: $\epsilon = 1/N = 1/9$).
Treatments 4 and 5 were variants of treatment 3, but 
some payoff parameters 
were changed in time to simulate varying environmental
conditions. In {\em treatment 5}, each player
additionally received an individual recommendation which alternative to choose.
\par
The higher changing rate in treatment 1 compared to treatment 2 can be 
understood as effect of an exploration rate $\nu_1$ required to find out which 
alternative performs better. It is also plausible that treatment 3 could further reduce
the changing rate: In the user equilibrium with $P_1(n_1) = P_2(n_2)$, every player knew that
he or she would {\em not} get {\em the same}, but a {\em reduced} payoff, if he or she would change the
decision. That explains why the new treatment 3 could reach 
a great adaptation performance, 
reflected by a very low standard deviation and almost optimal average payoffs. 
The behavioral changes induced by the treatments were not only observed on average, but for
every single individual (see Fig.~\ref{Fig4}). Moreover, even the
smallest individual cumulative payoff exceeded the highest one in treatment 1. Therefore, 
treatment 3's way of information presentation is much superior to the ones used today.

\section{Explaining the volatile decision dynamics}
\label{Volatile}

\begin{figure}[htbp]
\begin{center}
\begin{picture}(14,12.5)(0,6.7)
\put(-0.25,19.0){\includegraphics[width=6.05\unitlength, angle=-90]{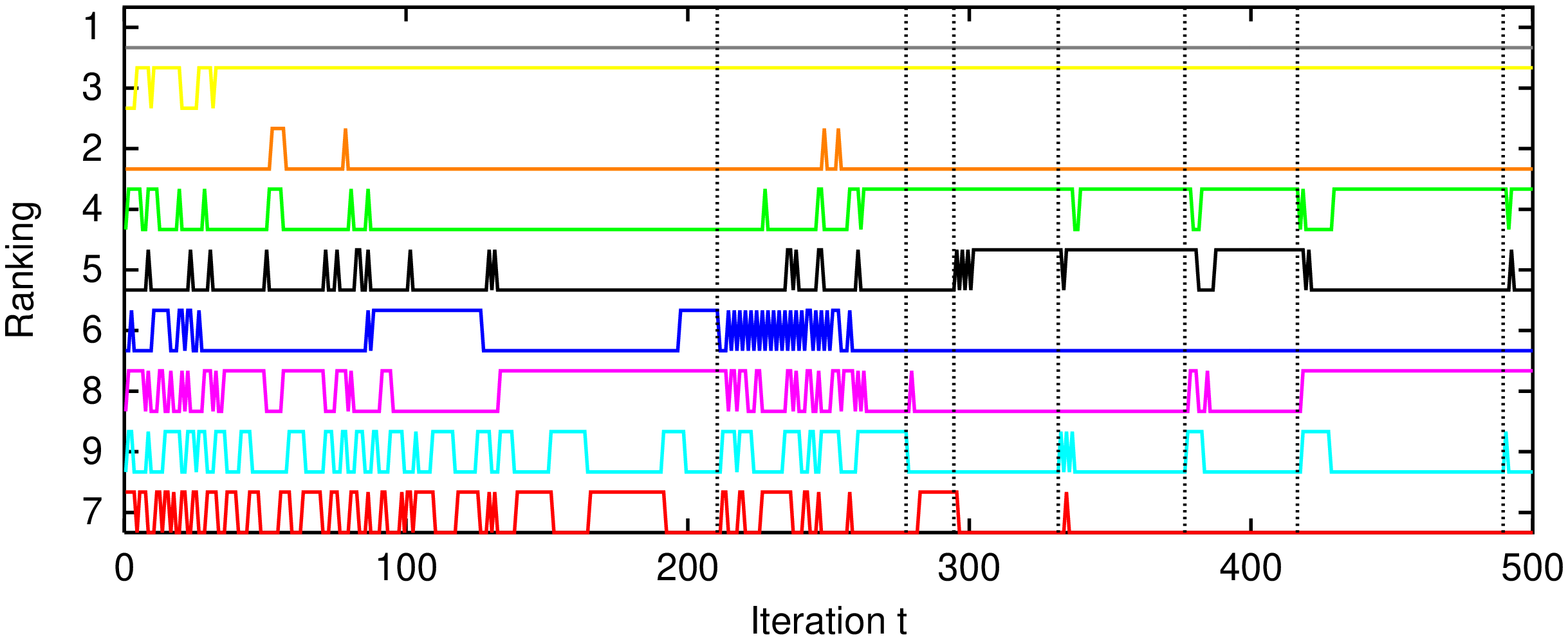}}
\put(-0.25,12.5){\includegraphics[width=6.05\unitlength, angle=-90]{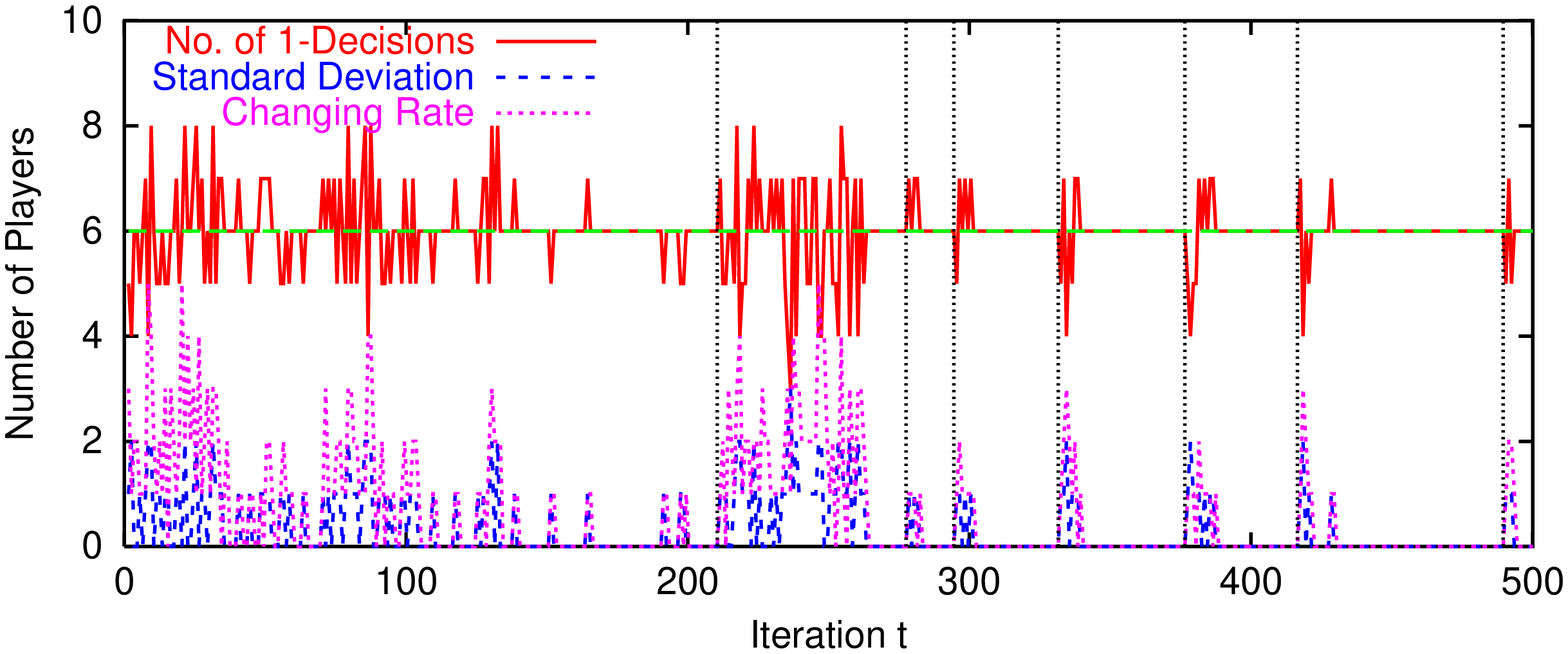}}
\put(-0.05,18.5){\sf\textbf{a}} 
\put(-0.05,12.0){\sf\textbf{b}} 
\end{picture}
\end{center}
\caption[]{Illustration of typical results for treatment 3 (which was here
the only treatment applied to the test persons, in contrast to Fig.~\ref{Fig3}).
{\sf\textbf{(a)}} Decisions of all 9 players. Players are displayed from the top to the bottom 
in the order of increasing
changing rate. Although the ranking of the cumulative payoff and the changing rate
are anticorrelated, the relation is not monotonic. Note that turbulent or volatile 
periods characterized by many decision changes are usually triggered by individual changes
after quiescent periods (dotted lines). {\sf\textbf{(b)}} 
The changing rate is mostly larger than
the (standard) deviation from the user equilibrium $n_1 = f_1^{\rm eq} N = 6$, 
indicating an overreaction in the
system.\label{Fig5}} 
\end{figure}

\begin{figure}[htbp]
\begin{center}
\includegraphics[width=4.3cm, angle=-90]{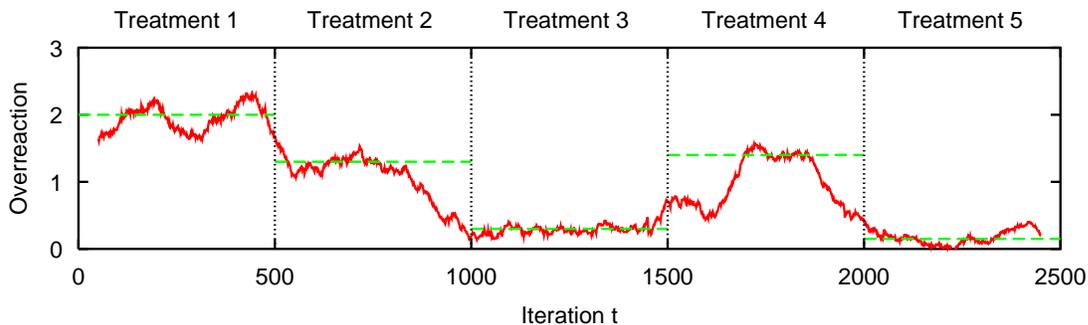}
\end{center}
\caption[]{Measured overreaction, i.e., difference between the actual 
number of decision changes  (the changing rate) and the required one 
(the standard deviation). The overreaction can be significantly influenced by the 
treatment, i.e. the way of information presentation. The minimum overreaction was
reached by treatment 5, i.e. user-specific recommendations.
\label{Fig6}}
\end{figure}

In this section, we will investigate why players changed their decision in the user equilibrium at all.
The reason for the pertaining changing behavior can be revealed by a more detailed
analysis of the individual decisions in treatment 3.
Figure~\ref{Fig5} shows some kind of intermittent behavior, 
i.e. quiescent periods without changes, followed by turbulent
periods with many changes. This is reminiscent of 
volatility clustering in stock market indices \cite{Stanley,Peinke,Lux}, 
where individuals also react to aggregate information reflecting all decisions 
(the trading transactions). 
Single players seem to change their decision to reach above-average
payoffs. In fact, although the cumulative individual payoff is anticorrelated 
with the average changing rate,
some players receive higher 
payoffs with larger changing rates than others. They 
profit from the overreaction in the system. 
Once the system is out of
equilibrium, all players respond in one way or another.
Typically, there are too many decision changes (see Figs.~\ref{Fig5} and \ref{Fig6}). 
The corresponding overcompensation, which had also been predicted by computer simulations
\cite{BarDin98,BenPaKa91,MahJa91,ArnPaLi91,Hall,WahBaKlSc00b},
gives rise to ``turbulent'' periods.  
\par
We should, however, note that the calm periods without decision changes tend to become
longer in the course of time. That is, after a very long time period the individuals learn 
not to change their behavior when the user equilibrium is reached. This is not only found in
Fig.~\ref{Fig5}, but also visible in Fig.~\ref{Fig3}{\sf c} after about 800 iterations. In larger
systems (with more participants) this transient period would take even longer, so that this
stabilization effect cannot be observed in experiments with less iterations or more
test persons.
\par
Finally, we should stress that other interpretations of the rather persistent decision changes
have been ruled out, for example, an unstable user equilibrium or a 
competition between the user optimum and
the system optimum. High changing rates also occur if
the user and system equilibrium agree, and if the payoff functions $P_1(n_1)$ and $P_2(n_2)$ 
are the same (i.e. $f_1^{\rm eq} = 1/2$).

\section{Decision and intermittency control by non-linear feedback based on guidance strategies}
\label{Control}
\begin{figure}[htbp]
\begin{center}
\begin{picture}(14,12)(0,7.2) 
\put(0.1,19.0){\includegraphics[width=6.05\unitlength, angle=-90]{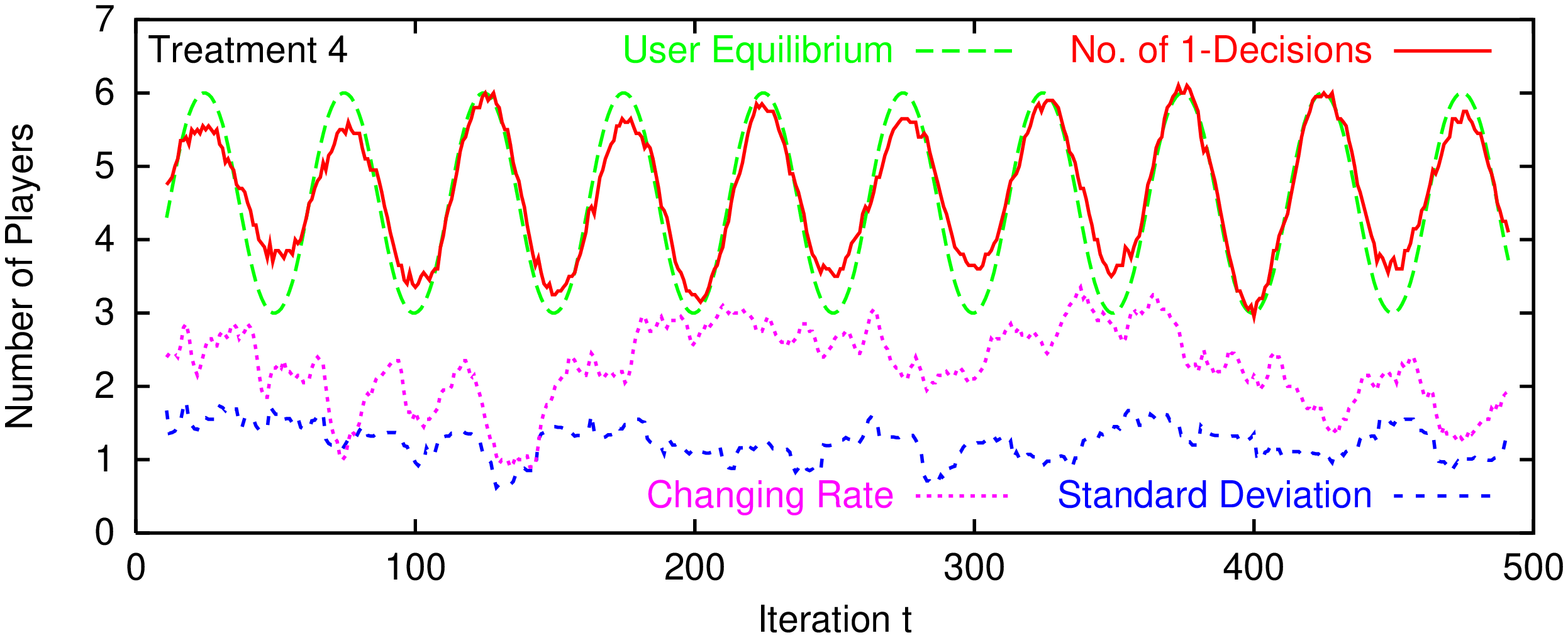}}
\put(0.1,13.0){\includegraphics[width=6.05\unitlength, angle=-90]{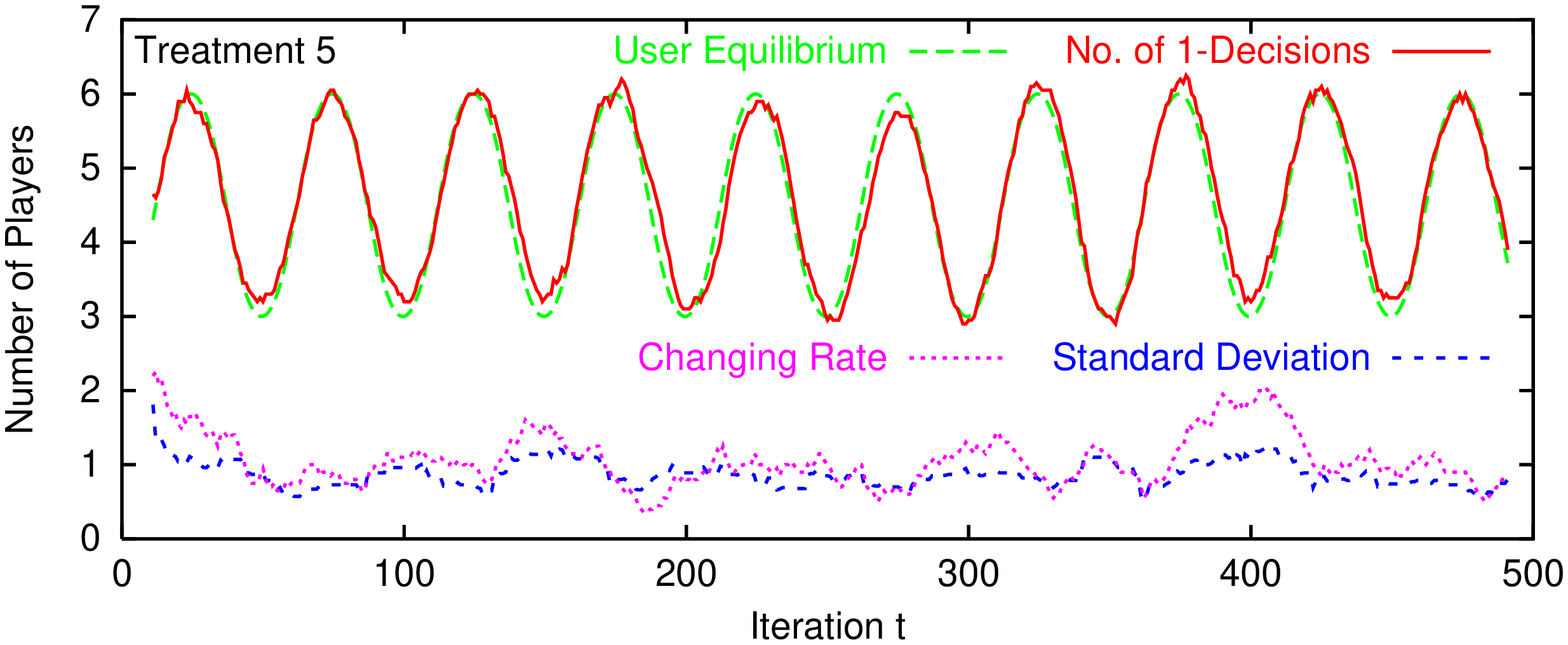}} 
\put(0.05,18.5){\sf\textbf{a}}
\put(0.1,12.5){\sf\textbf{b}} 
\end{picture}
\end{center}
\caption[]{Representative examples for  {\sf\textbf{(a)}}
treatment 4 and  {\sf\textbf{(b)}} treatment 5. 
The displayed curves are moving time-averages over 20 iterations.
Compared to treatment 4, the user-specific recommendations in treatment 5 
(assuming $C_M = C_S = 1$, $R_1 = 0$, 
$R_2= \max([f_1^{\rm eq}(t+1)N - n_1(t)+B(t+1)]/n_2(t),1)$,  $I_1 = I_2 = 1$) 
could increase the group adaptability to the 
user equilibrium a lot, even if they had a systematic or random 
bias $B$ (see Fig.~\ref{Fig8}{\sf a}).
The standard deviation was reduced considerably and the changing rate even more.
\label{Fig7}}
\end{figure}

To avoid overreaction, 
in treatment 5 we have recommended a number \mbox{$f_1^{\rm eq}(t+1)N - n_1(t)$}
of players to change their decision and the other ones to keep it. These user-specific 
recommendations helped the players to reach the 
smallest overreaction of all treatments (see Fig.~\ref{Fig6})
and a very low standard deviation, although the
payoffs were changing in time (see Fig.~\ref{Fig7}). 
Treatment 4 shows how the group performance was affected
by the time-dependent user equilibrium: Even without re\-commendations,
the group managed to adapt to the changing conditions surprisingly well, but the 
standard deviation and changing rate were approximately as high as in treatment 2
(see Fig.~\ref{Fig3}). This adaptability (the collective ``group intelligence'') is based on complementary
responses (direct and contrary ones \cite{Selten}, ``movers'' and ``stayers'', cf. Fig.~\ref{Fig4}).
That is, if some players do not react to the changing conditions, others will
take the chance to earn additional payoff. This experimentally supports the
behavior assumed in the theory of efficient markets, but here the efficiency is limited
by overreaction.
\par
\begin{figure}[htbp]
\begin{center}
\begin{picture}(14,6.2)(0,0.5)
\put(-0.25,6.7){\includegraphics[width=6.45\unitlength, angle=-90]{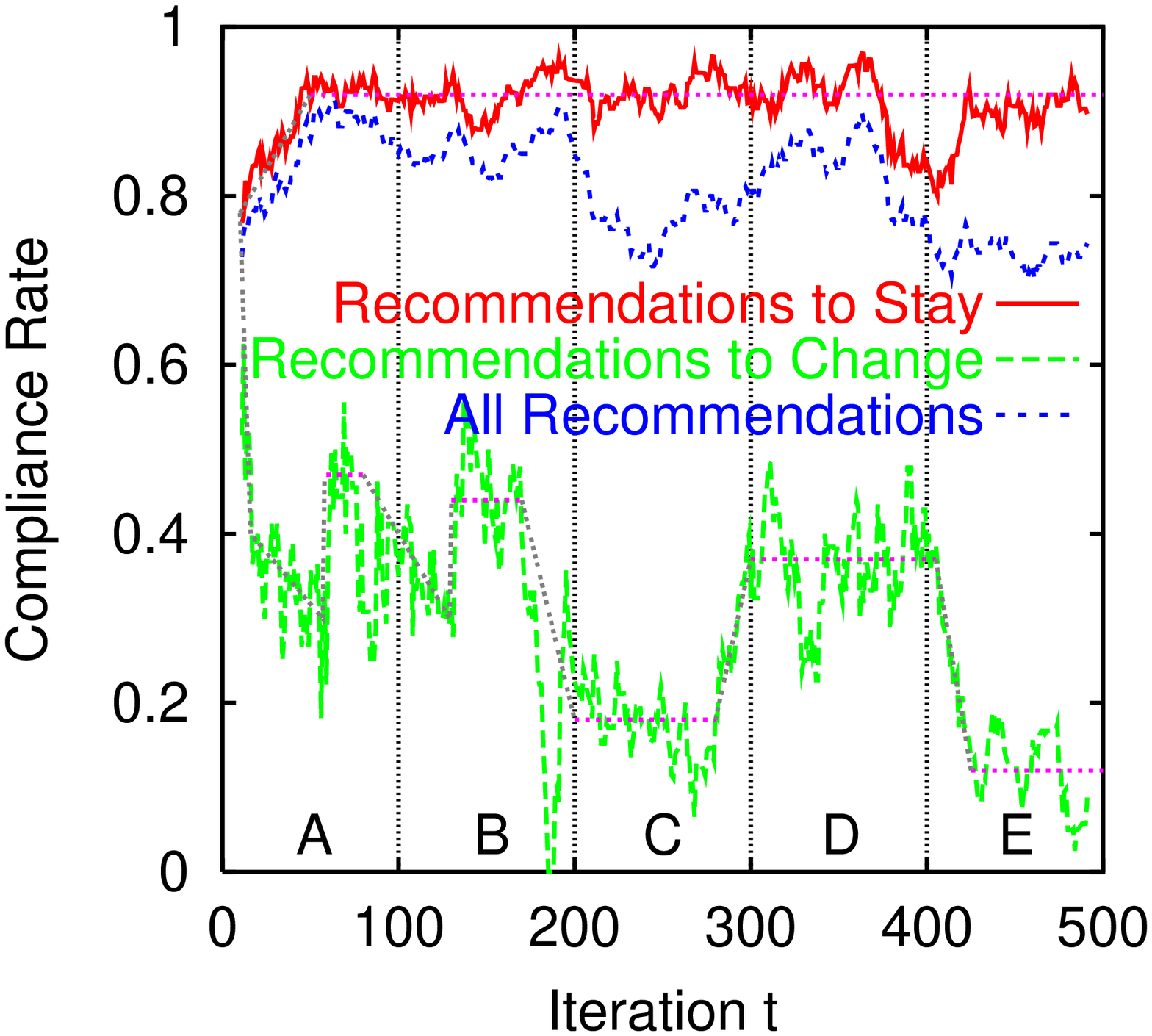}}
\put(7.15,6.7){\includegraphics[width=6.45\unitlength, angle=-90]{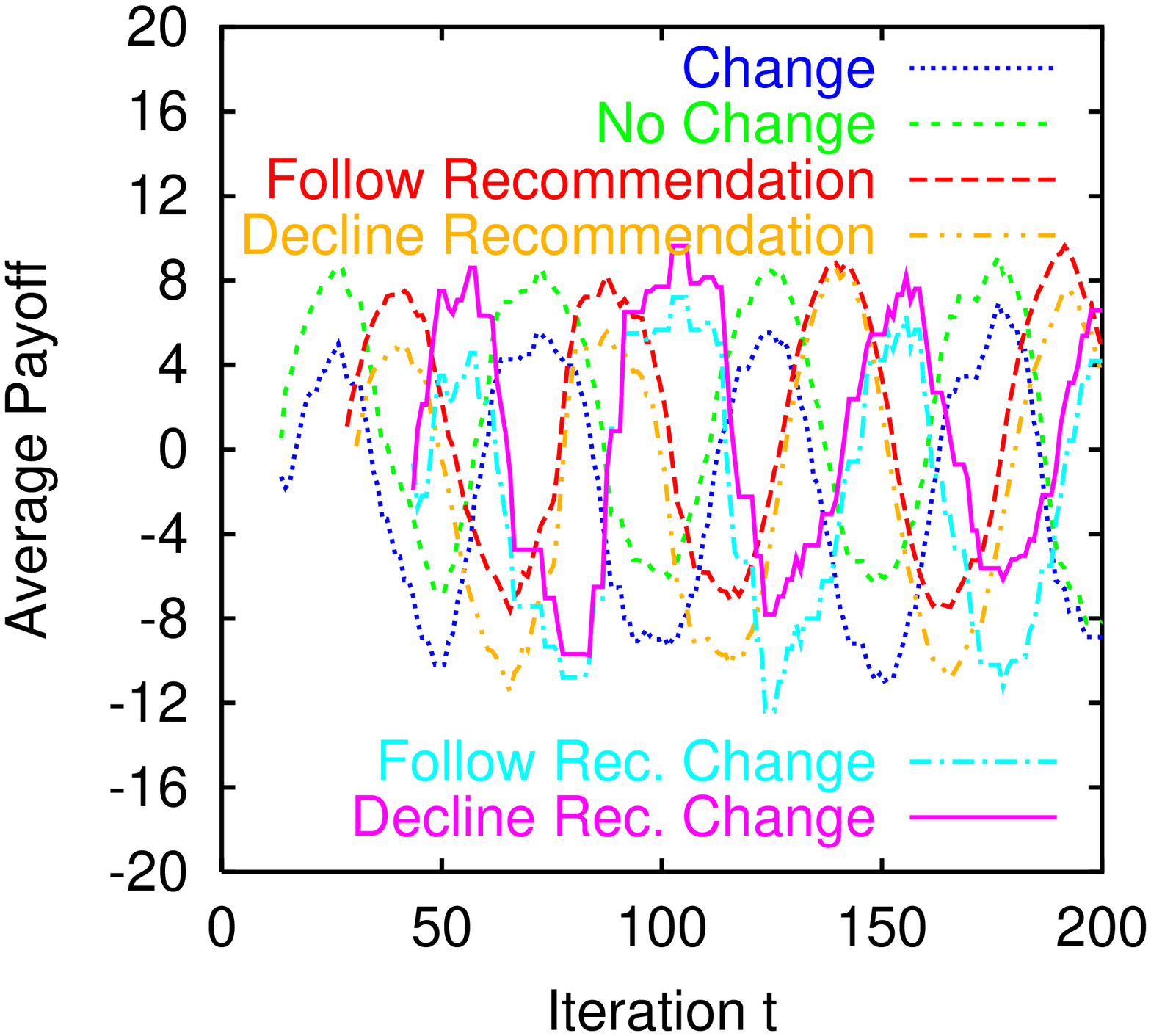}}
\put(-0.05,6.2){\sf\textbf{a}} 
\put(7.5,6.2){\sf\textbf{b}} 
\end{picture}
\end{center}
\caption[]{{\sf\textbf{(a)}} 
In treatment 5, the compliance to recommendations to change 
dropped considerably below the compliance to recommendations to stay. 
The compliance to changing recommendations was very sensitive to the degree of their reliability,
i.e. participants followed recommendations just as much 
as they helped them to reach the user equilibrium (so that the bias $B$ did not affect the
small deviation from it, see Fig.~\ref{Fig7}{\sf b}).
While during time interval {\sf{A}}, the recommendations would have been perfect, if all 
players had followed them, in time interval {\sf{B}} the user equilibrium was overestimated
by $B=+1$, in {\sf{C}} it was underestimated by $B=-2$, in {\sf{D}} it was randomly 
over- or underestimated by $B=\pm 1$, and in {\sf{E}} by $B=\pm 2$. Obviously,
a random error is more serious than a systematic one of the same amplitude. 
Dotted non-vertical lines illustrate the estimated compliance levels during 
the transient periods and afterwards (horizontal dotted lines).
{\sf\textbf{(b)}}  The average payoffs varied largely with the decision behavior. 
Players who changed their decision got significantly lower payoffs on average than
those who kept their previous decision. Even recommendations could
not overcome this difference: It stayed profitable not to change,
although it was generally better to follow recommendations than to refuse them. 
For illustrative reasons, the third and fourth line were shifted by 15, while the
fifth and sixth line were shifted by 30 iterations.\label{Fig8}}
\end{figure}

In most experiments, we found a constant and high compliance $C_S(t) \approx 0.92$
with recommendations to stay, but the compliance $C_M(t)$ with recommendations
to change (to `move') \cite{Chen,Kuehne,Kraan,Srinivasan} 
turned out to vary in time. It decreased with the 
reliability of the recommendations 
(see Fig.~\ref{Fig8}{\sf a}), which again dropped with the compliance. 
\par
Based on this knowledge, we have developed a model, how the competition for limited resources 
(such as road capacity) could be {\em optimally} guided by means of information services. Let us 
assume we had $n_1(t)$ 1-decisions at time $t$, but the optimal number of 1-decision at time
$t+1$ is calculated to be $f_1^{\rm eq}(t+1)N \ge n_1(t)$. Our aim is to balance the
deviation $f_1^{\rm eq}(t+1)N - n_1(t) \ge 0$ by the expected net number 
\begin{equation}
\langle \Delta n_1(t+1) \rangle =
\langle n_1(t+1) - n_1(t) \rangle = \langle n_1(t+1) \rangle - n_1(t) 
\end{equation}
of transitions from decision 2 to decision 1, i.e.
$f_1^{\rm eq}(t+1)N - n_1(t) = \langle \Delta n_1(t+1) \rangle$. In the case
$f_1^{\rm eq}(t+1)N - n_1(t) < 0$, 
indices 1 and 2 have to be interchanged.
\par
Let us assume we give
recommendations to fractions $I_1(t)$ and $I_2(t)$ of players who had chosen
decision 1 and 2, respectively. The fraction of changing recommendations to
previous 1-choosers shall be denoted by $R_1(t)$, and for previous 2-choosers by
$R_2(t)$. Correspondingly, fractions of $[1-R_1(t)]$ and $[1-R_2(t)]$ 
receive a recommendation to stick to the  previous decision. 
Moreover, $[1-C_M(t)]$ is the {\em refusal probability} of recommendations to change, while
 $[1-C_S(t)]$ is the refusal probability of recommendations to stay. 
Finally, we denote the spontaneous transition probability 
from decision 1 to 2 by $p_a(2|1,n_1;t)$ and the inverse transition probability by
$p_a(1|2,n_1;t)$, in case a player does not receive any recommendation. This happens with
probabilities $[1-I_1(t)]$ and $[1-I_2(t)]$, respectively. Both transition probabilities
$p_a(2|1,n_1;t)$ and $p_a(1|2,n_1;t)$
are functions of the number $n_1(t) = N -n_2(t)$ of previous 1-decisions. The index $a$
allows us to reflect different strategies  or characters of players. 
The fraction of players pursuing strategy $a$ is then denoted by $F_a(t)$.
Applying methods summarized in Ref. \cite{Diss},
the expected change $\langle \Delta n_1(t+1) \rangle$
of $n_1$ is given by the balance equation
\begin{eqnarray}
\langle \Delta n_1(t+1) \rangle &=& \sum_a p_a(1|2,n_1;t) F_a(t) [1- I_2(t)] n_2(t) \nonumber \\
                                  &-& \sum_a p_a(2|1,n_1;t) F_a(t) [1-I_1(t)] n_1(t)  \nonumber \\
                &+& \sum_a \{ C_M^a(t) R_2(t) 
                                    + [1- C_S^a(t)] [1-R_2(t)]\} F_a(t) I_2(t) n_2(t) \nonumber \\
                                  &-& \sum_a \{ C_M^a(t) R_1(t) 
                                     + [1-C_S^a(t)]  [1-R_1(t)]\} F_a(t) I_1(t) n_1(t) \, .\quad
\label{balance}
\end{eqnarray}
Together with the requirement
\begin{equation}
 \langle \Delta n_1(t+1) \rangle = f_1^{\rm eq}(t+1)N - n_1(t) \, ,
\label{bal}
\end{equation}
this equation defines, with respect to the number $n_1$ of previous 1-decisions,  
a {\em non-linear feedback} or {\em control strategy}.
\par
\begin{figure}[htbp]
\begin{center}
\begin{picture}(14,6.2)
\put(-0.55,6.2){\includegraphics[width=6.45\unitlength, angle=-90]{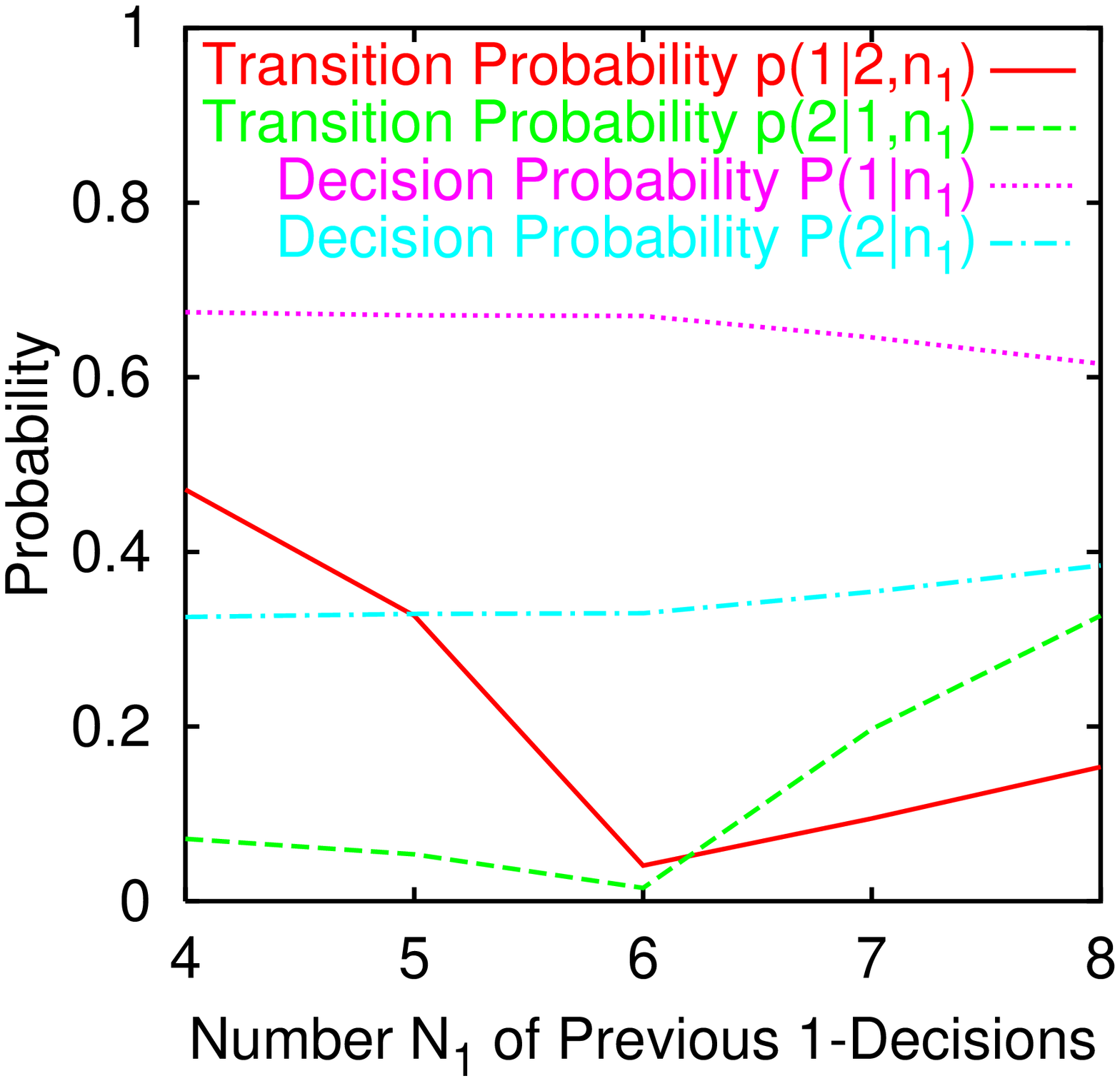}}
\put(6.75,6.2){\includegraphics[width=6.45\unitlength, angle=-90]{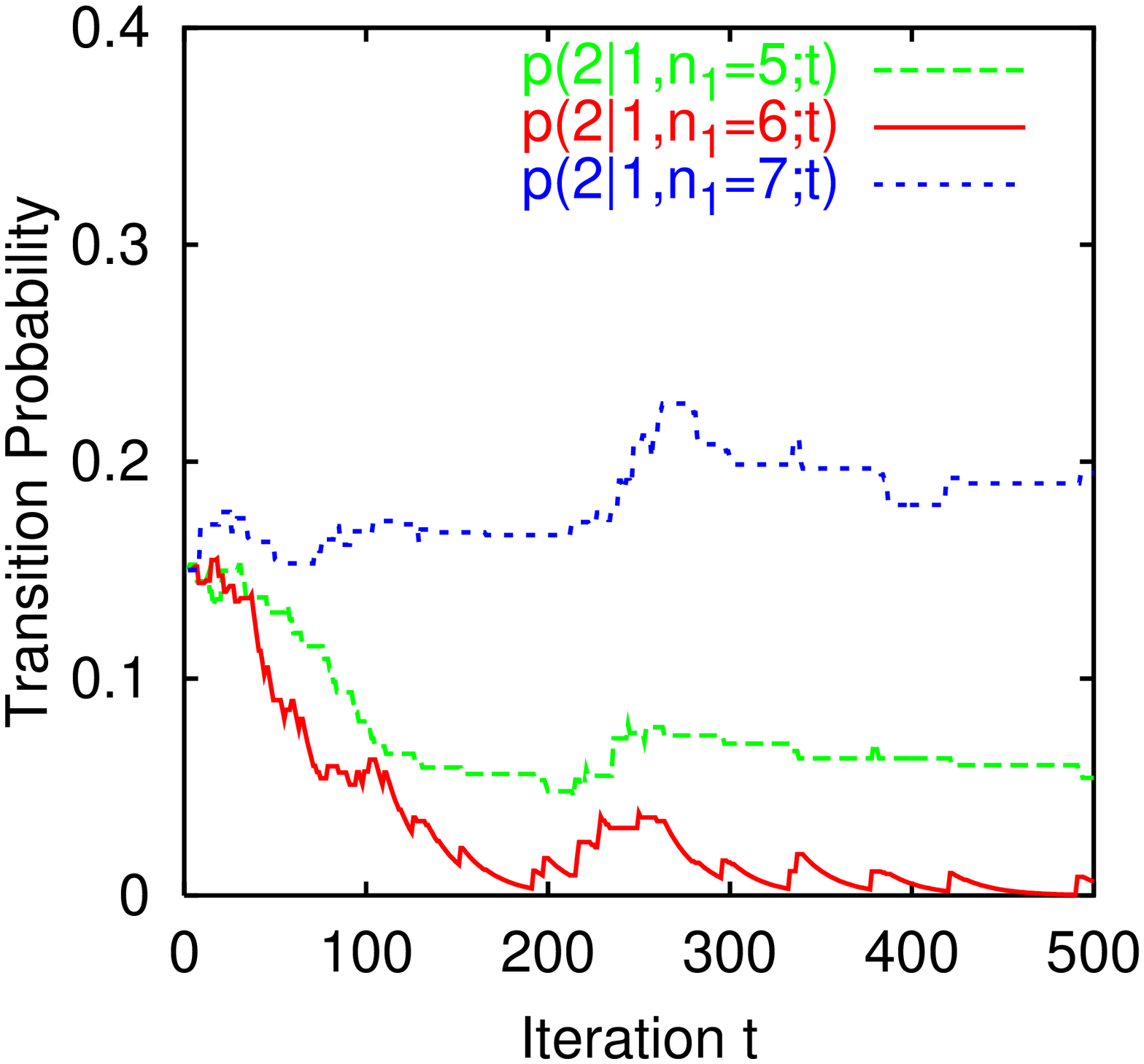}}
\put(-0.05,5.7){\sf\textbf{a}} 
\put(7.2,5.7){\sf\textbf{b}} 
\end{picture}
\end{center}
\caption[]{Illustration of decision distributions
$P(i|n_1)$ and transition probabilities $p(i'|i,n_1;t)$ measured in treatment 3. 
{\sf\textbf{(a)}} The probability $P(1|n_1)$ to choose alternative 1 was approximately
2/3, independently of the number $n_1$ of players who had previously chosen alternative 1.
The probability $P(2|n_1)$ to choose alternative 2, given that $n_1$ players had chosen alternative 1,
was always about 1/3. In contrast, the transition probability $p(1|2,n_1)$ describing decision changes from 
alternative 2 to 1 did depend on the number $n_1$ of players who had chosen decision 1.
The same was true for the  inverse transition probability 
$p(2|1,n_1)$ from decision 1 to decision 2. 
Remarkably enough, these transition probabilities are 
not monotonically increasing with the payoff or the expected payoff gain, as they
do not monotonically increase with $n_1$. 
Instead, the probability to change the decision shows a minimum at the user equilibrium
$n_1 = f_1^{\rm eq} N =  6$. {\sf\textbf{(b)}} The reason for the different transition probabilities
is an adaptation process in which the participants learn to take fewer changing decisions, when 
the user equilibrium is reached or close by, but more, when the user equilibrium
is far away. (The curves were exponentially smoothed with $\alpha = 0.05$.)
\label{Fig9}}
\end{figure}

Note that, for Eq.~(\ref{balance}),  it was not necessary to distinguish different characters $a$. 
We have, therefore, evaluated the overall transition probabilities 
\begin{equation}
 p(1|2,n_1;t) = \sum_a p_a(1|2,n_1;t)F_a(t)
\quad \mbox{and} \quad p(2|1,n_1;t) = \sum_a p_a(2|1,n_1;t) F_a(t) \, .
\end{equation}
According to classical decision theories \cite{Diss,Ortuzar,Benetal,Lerman}, 
we would expect that the transition
probabilities $p_a(1|2,n_1;t)$ and $p(1|2,n_1;t)$ 
should be monotonically increasing functions of the payoff 
$P_1(n_1(t))$, the payoff difference $P_1(n_1(t)) - P_2(N-n_1(t))$, the potential
payoff $P_1(n_1(t)+\epsilon N)$, or the potential payoff gain 
$P_1(n_1(t)+\epsilon N) - P_2(N-n_1(t))$.
All these quantities vary linearly with $n_1$, so that $p(1|2,n_1;t)$ should be
a monotonic function of $n_1(t)$. A similar thing should apply to $p(2|1,n_1;t)$.
Instead, the experimental data point to transition probabilities with 
a {\em minimum} at the user equilibrium (see Fig.~\ref{Fig9}{\sf{a}}).
That is, the players stick to a certain
alternative for a longer time, when the system is close to the user equilibrium.
This is a result of learning \cite{Nakayama} (see also Refs.
\cite{learn2,learn3,learn4,coordination,adaptive}). 
In fact, we find a gradual change of the transition
probabilities in time (see Fig.~\ref{Fig9}{\sf{b}}).  
The corresponding ``learning curves'' reflect the players' adaptation to the user equilibrium.
\par
After the experimental determination of the transition probabilities $p(2|1,n_1;t)$, $p(1|2,n_1;t)$ 
and specification of the overall compliance probabilities 
\begin{equation}
C_M(t) = \sum_a C_M^a(t)F_a(t)\, , \qquad C_S(t) = \sum_a C_S^a(t) F_a(t) \, ,
\end{equation} 
we can guide the decision behavior in the system
via the levels $I_i(t)$ of information dissemination and the
fractions $R_i(t)$ of recommendations to change ($i\in \{1,2\}$). 
These four degrees of freedom allow us to apply a variety of guidance strategies depending
on the respective information medium. For example, 
a guidance by radio news is limited by the
fact that  $I_1(t) = I_2(t)$ is given by the average percentage of radio users.  Therefore,
equations (\ref{balance}) and (\ref{bal}) cannot always be solved by variation of the fractions of
changing recommendations $R_i(t)$.
User-specific services have much higher guidance 
potentials and could, for example, be transmitted via SMS.
Among the different guidance strategies fulfilling equations (\ref{balance}) and (\ref{bal}), 
the one with the minimal statistical variance will be the best. However, it would
already improve the present situation to inform {everyone} about the {\em fractions} $R_i(t)$ of 
participants who should change their decision, as users can learn to respond with varying 
probabilities (see Fig.~\ref{Fig9}). 
\par
The outlined guidance 
strategy could, of course, also be applied to reach the system optimum rather than the
user optimum. The values of $\Delta n_1(t+1)$ would just be different. Note, however,
that the users would soon recognize that this guidance is not suitable to reach the user optimum.
Consequently, the compliance probabilities $C_j(t)$ with $j\in \{M,S\}$ would gradually go down, 
which would affect the potentials and reliability
of the guidance system. This problem can only be solved by a suitable modification of the 
payoff functions, adapting the user optimum to the system optimum.
\par
In practical applications, we would 
determine the time-dependent compliance probabilities $C_j(t)$ (and the
transition probabilities)  on-line with an exponential smoothing procedure according to 
\begin{equation}
 C_j(t+1) = \alpha C'_j(t) + (1-\alpha) C_j(t) \quad \mbox{with} \quad \alpha \approx 0.1\, ,
\end{equation}
where $C'_j(t)$ is the percentage of participants 
who have followed their recommendation at time $t$. 
As the average payoff for decision changes is normally lower than for staying with 
the previous decision (see Figs.~\ref{Fig8}{\sf b} and \ref{Fig3}{\sf{d}}), a
high compliance probability $C_M$ is hard to achieve. That is, individuals who follow 
recommendations to change normally pay for reaching the user equilibrium
(because of the overreaction in the system).
Hence, there are no good preconditions to charge the players for recommendations, 
as we did in another treatment. 
Consequently, only a few players requested recommendations, which reduced their reliability,
so that the overall performance of the system went down.

\section{Master equation description of iterated decisions}
\label{MASTEReq}
The stochastic description of decisions that are taken at discrete time steps (e.g. on a day-to-day
basis) is possible by means of the time-discrete master equation \cite{REVIEW}
\begin{equation}
 P(\vec{n},t+\Delta t) = \sum_{\vec{n}'}
 P(\vec{n},t+\Delta t|\vec{n}',t) P(\vec{n}',t) 
\label{markov}
\end{equation}
with $\Delta t = 1$ unit time. 
Herein, $P(\vec{n},t)$ denotes the occurence probability 
of the {\em configuration} $\vec{n} = (n_1,n_2)$ at time $t$. This vector
comprises the {\em occupation numbers} $n_i$ and reflects the decision distribution
in the system. As the number of individuals changing
to the other alternative is given by a {\em binomial distribution}, we obtain
the following expression for the {\em configurational transition probability}:
\begin{eqnarray}
\hspace*{-6mm} & & P\big((n_1,n_2),t+1 \, \big| \, (n_1-\Delta n_1,n_2+\Delta n_1),t\big) \nonumber \\
\hspace*{-6mm} && \quad = \!\!\! \sum_{k=0}^{\min(n_1-\Delta n_1,n_2)}
\left(
\begin{array}{c}
n_2+\Delta n_1\\
\Delta n_1 + k
\end{array}
\right) 
p(1|2,n_1-\Delta n_1;t)^{\Delta n_1 + k} \big[ 1 - p(1|2,n_1-\Delta n_1;t) \big]^{n_2 - k}
\nonumber \\
\hspace*{-6mm}& & \quad \qquad \times
\left(
\begin{array}{c}
n_1-\Delta n_1\\
k
\end{array}
\right) 
p(2|1,n_1-\Delta n_1;t)^{k} \big[ 1 - p(2|1,n_1-\Delta n_1;t) \big]^{n_1-\Delta n_1 - k} \, .
\label{binom}
\end{eqnarray}
This formula sums up the probabilities that $\Delta n_1 + k$ of $n_2+\Delta n_1$
previous 2-choosers change independently to alternative 1 with probability 
$p(1|2,n_1-\Delta n_1;t)$, while $k$ of the $n_1-\Delta n_1$ previous 1-choosers
change to alternative 2 with probability  $p(2|1,n_1-\Delta n_1;t)$, so that the
net number of changes is $\Delta n_1$.
If $\Delta n_1 < 0$, the roles of alternatives 1 and 2 have to be interchanged. 
Formulas (\ref{markov}) and (\ref{binom}) would look even more compicated, if we distinguished
several characters $a$. We would, then, have to replace the binomial distributions by
multinomial ones.
\par
The potential use of Eq. (\ref{binom}) is the calculation of the statistical
variation of the decision distribution or, equivalently, the number $n_1$ of 1-choosers. 
It also allows one to determine the variance, which the optimal guidance strategy should
minimize in favour of reliable recommendations.

\section{Summary and Outlook}

In this contribution, we have discovered that the dynamics of iterated decisions based on
aggregate information is intermittent. In order to control intermittency, we
have developed a stochastic description and a non-linear feedback mechanism.
That is, the application of several physical concepts and methods allowed us to
gain a detailled understanding of decision dynamics, which is required for practical applications. 
\par
In more detail,
we have explored different and identified superior ways of information presentation that facilitate
to guide user decisions in the spirit of higher payoffs. By far the
least standard deviations from the user equilibrium could be reached by presenting
the own payoff and the potential payoff, if the respective participant 
(or a certain fraction of players)
had additionally chosen the other alternative. Interestingly, the decision dynamics was found to
be intermittent similar to the volatility clustering in 
stock markets,  where individuals also react to aggregate information. This results from 
the desire to reach above-average payoffs, 
combined with the immanent overreaction in the system. We have also demonstrated that
payoff losses due to a volatile decision dynamics  (e.g., excess travel times)
can be reduced via user-specific recommendations by a factor of three or more.  
Such kinds of results will be applied to the route guidance on German highways
(see, for example, the project SURVIVE conducted by Nobel prize winner Reinhard 
Selten and Michael Schreckenberg). 
Optimal recommendations to reach the user equilibrium follow directly from 
the derived balance equations (\ref{balance}) and (\ref{bal}) for decision changes based on empirical 
transition and compliance probabilities. The quantification of the transition probabilities needs 
a novel stochastic description of the decision behavior, which is not
just driven by the potential (gains in) payoffs, in contrast to intuition and established models. 
To understand these findings, one has to take into account individual learning. 
\par
Obviously, it requires both, theoretical and experimental efforts to get ahead 
in decision theory. In a decade from now, the theory of  ``elementary'' human interactions
will probably have been developed to a degree that allows one to systematically derive
social patterns and economic dynamics on this ground in a similar way 
as the structure, properties, and dynamics of matter 
have been derived from elementary physical interactions. This will not only yield a
deeper understanding of socio-economic systems, but also help to more efficiently
distribute scarce resources such as road capacities, time, space, money, energy, goods,
or our natural environment. One day, similar guidance strategies as the ones
suggested above may help politicians and managers to stabilize economic markets,
to increase average {\em and} individual profits, and to decrease the unemployment rate.
Physics can contribute to this goal, in particular with the methods developed in 
the fields of non-linear dynamics and statistical physics.

{\bf Acknowledgment:} The authors are grateful to the ALTANA-Quandt
foundation for financial support, to Tilo Grigat for preparing
some of the illustrations, and to the test persons.

\end{document}